\documentclass{aa}
\usepackage{graphicx}
\usepackage{txfonts}
\usepackage{latexsym}
\usepackage{times,amssymb,lscape,verbatim}
\usepackage{natbib}
\usepackage[mathcal]{euscript}
\usepackage{mathrsfs}

\begin{document}
	\title{SMA observations of young disks: separating envelope, disk, and stellar masses in class I YSOs}
   \titlerunning{Separating the masses in Class~I YSOs}

   \author{Dave Lommen\inst{1}
	\and Jes K. J{\o}rgensen\inst{2,3}
	\and Ewine F. van Dishoeck\inst{1}
	\and Antonio Crapsi\inst{1,4}
	     }

   \offprints{Dave Lommen, \\ \email{dave@strw.leidenuniv.nl}}

   \institute{Leiden Observatory, Leiden University, P.O. Box 9513, 2300 RA Leiden, The Netherlands
	\and Harvard-Smithsonian Center for Astrophysics, 60 Garden Street, Cambridge, MA 02138, USA
	\and Argelander-Institut f\"{u}r Astronomie, University of Bonn, Auf dem H\"{u}gel 71, 53121 Bonn, Germany
	\and Observatorio Astron\'{o}mico Nacional (IGN), Alfonso XII, 3, 28014 Madrid, Spain
             }

   \date{Received ...; accepted ...}

 \abstract 
  {Young stars are born with envelopes, which in the early stages obscure the central (proto)star and circumstellar disk.
    In the Class I stage, the disks are still young, but the envelopes are largely dispersed. This makes the Class I sources ideal targets for studies of
    the early stages of disks.}
 {We aim to determine the masses of the envelopes, disks, and central stars of young stellar objects (YSOs) in the Class I stage.}
  {We observed the embedded Class I objects IRS~63 and Elias~29 in the $\rho$~Ophiuchi star-forming region with the Submillimeter Array (SMA) at 1.1~mm.}
{IRS~63 and Elias~29 are both clearly detected in the continuum, with peak fluxes of 459 and 47~mJy/beam, respectively. The continuum emission
toward Elias~29 is clearly resolved, whereas IRS~63 is consistent with a point source down to a scale of 3$\arcsec$ (400~AU). The SMA data are combined with single-dish data, and
both disk masses of 0.055 and $\le 0.007$~M$_\odot$ and envelope masses of 0.058 and $\le 0.058$~M$_\odot$ are empirically determined for IRS~63 and Elias~29, respectively. The
disk+envelope systems are modelled with the axisymmetric radiative-transfer code RADMC, yielding disk and envelope masses that differ from the empirical results by factors of a
few. HCO$^+$ $J$ = 3--2 is detected toward both sources, HCN $J$ = 3--2 
is not. 
The HCO$^+$ position-velocity diagrams are indicative of Keplerian rotation and allow an estimate of the mass of the central stars.
For a fiducial inclination of 30$^\circ$, we find stellar masses of $0.37 \pm 0.13$ for IRS~63 and $2.5 \pm 0.6$~M$_\odot$ for Elias~29.}
{The sensitivity and spatial resolution of the SMA at 1.1~mm allow a good separation of the disks around Class~I YSOs from their circumstellar envelopes and environments,
and the spectral resolution makes it possible to resolve their dynamical structure and estimate the masses of the central stars.
The ratios of the envelope and disk masses $M_{\rm env}/M_{\rm disk}$ are found to be 0.2 for IRS~63 and 6 for Elias~29. 
This is lower than the values for Class~0 sources, which have $M_{\rm env}/M_{\rm disk} \ge 10$, suggesting that this ratio is a tracer of
the evolutionary stage of a YSO.}


   \maketitle
%

\section{Introduction}\label{sect: introduction}

	Low- and intermediate-mass stars are formed from the graviational collapse of molecular cloud cores. In the earliest stages, the
	newly-formed protostar remains embedded in the remnants of this core, a cold envelope of dust and gas, which is gradually accreted by the young star 
	(e.g., Shu 1977). Due to
	the angular momentum initially present in the core, most of the envelope material does not fall directly onto the central protostar but is piled-up in a
	circumstellar disk (e.g., Terebey et al. 1984). Understanding this interplay between star, disk, and envelope is crucial in order to be
	able to relate the initial conditions of star formation such as the mass of the protostellar core to the end-product -- namely the properties of the
	young star, and the mass and thus potential of the disk for forming planets. Some of the key questions include:
	Where does most of the mass reside at a given time? Will all the mass that 
	is seen in prestellar cores or in the envelopes around deeply embedded sources end up in the star, or will a large fraction be dispersed from the 
	system? How do the masses of the circumstellar envelope, the disk, and that of the central star evolve over time, and how long does it take for the 
	circumstellar matter to be accreted onto the star?

	Young stellar objects (YSOs) are usually classified according to their slopes in the infrared (IR) wavelength regime. Originally, the LW classification \citep{lada:1984,
	lada:1987} ran from the embedded Class I, via the optically visible Class II or classical T Tauri stars, to the Class III spectral energy distributions of post-T 
	Tauri stars.
	Later, the Class 0 stage was added to this classification \citep[see, e.g.,][]{andre:1993}, where the Class 0 sources are 
	distinguished from the Class I sources through their high relative luminosity at submillimetre (submm) wavelengths.
	This classification roughly reflects
	the evolutionary stage of the YSOs under consideration. The most deeply embedded 
	Class 0 sources are thought to evolve through the Class I stage while
	dissipating their circumstellar envelopes. Eventually they become optically visible as pre-main sequence T Tauri stars with circumstellar disks.
	
	The Class 0 and Class II YSOs have been studied quite extensively with high-resolution (sub)millimetre
	interferometers \citep[e.g.,][and references therein]{jorgensen:2007, andrews:2007}. Studies of deeply embedded Class~0 YSOs have shown that 
	circumstellar disks are	formed early \citep{harvey:2003, jorgensen:2004}, but in these systems, 
	it is difficult to separate the emission from the disk from that of the envelope.
	The Class~I sources, in which a large part of the original circumstellar envelopes has been dissipated, are ideal to study young disks in 
	YSOs. Interferometric studies of Class I objects have so far been largely limited
	to studies at around 3~mm \citep[e.g.,][]{ohashi:1997, hogerheijde:1997a, hogerheijde:1998, looney:2000}.
	The Submillimeter Array (SMA) allows observations around 1~mm, where the thermal dust continuum emission is an order of magnitude stronger
	than at 3~mm. Also at these shorter wavelengths it is possible to detect the higher rotational transitions of the molecules that trace the dense 
	gas in the disks and inner envelopes of the young systems, rather than lower-density extended envelope emission.
	
	We here present SMA observations of two Class I objects that appear to be in an evolutionary stage
	where the kinematics of the circumstellar material are no longer dominated by infall. \object{IRS 63} (\object{WLY 2-63}, 
	\object{GWAYL 4}) and \object{Elias 29} (\object{Elia 2-29}, \object{WLY 1-7}) are located in the $\rho$~Ophiuchi cloud, taken to be at a
	distance D~$= 125 \pm 25$~pc \citep{degeus:1989}. Spitzer photometry from the ``Cores to Disks'' Legacy program \citep{evans:2003} was used to
	determine the source's infrared colours and their bolometric luminosities and temperatures. IRS~63 and Elias~29 have similar bolometric 
	temperatures, $T_{\rm bol} = 351$~K and 391~K, respectively, which places them in the LW Class~I regime. The bolometric luminosities were
	calculated to be $L_{\rm bol} = 0.79$~L$_\odot$ for IRS~63 and 13.6~L$_\odot$ for Elias~29. According to their infrared slopes of
	$\alpha_{2-24 \mu m} = 0.15$ and 0.42, IRS~63 is the slightly more evolved, falling into the ``flat-spectrum'' class that separates the Class~I
	from the Class~II sources \citep{greene:1994}.
	The values for $\alpha_{\rm IR}$ quoted here are from un-dereddened observations.
	Correction for the extinction toward the $\rho$~Ophiuchi star-forming region \citep[e.g.,][]{flaherty:2007} 
	would result in lower values for $\alpha_{\rm IR}$; in other words, the sources may be slightly more evolved than the raw values of 
	$\alpha_{\rm IR}$ suggest.
	In terms of environment, the two sources are quite each other's opposites. The SCUBA 850~$\mu$m map \citep{johnstone:2004} from the 
	COMPLETE survey \citep{ridge:2006} shows that IRS~63 is an isolated compact source. Elias~29, on the other hand, is located in a dense ridge of 
	molecular material, which contains several more YSOs. It is likely that these YSOs were formed from condensations in this molecular ridge.
	
	We present the masses of all main components of Class~I YSOs, i.e., the central star, the disk, and the envelope, for the first time. 
	IRS~63 and Elias~29 are the
	first two sources in a larger survey of Class~I objects studied with the SMA at 1.1~mm. These observations complement the survey of Class~0 sources in the
	PROSAC programme \citep{jorgensen:2007} and will allow us to trace the similarities and differences of these evolving protostars.
	The results of the complete campaign will be presented in a
	future paper. In \S\ref{sect: observations}, the observations are presented, and the results and implications are discussed in 
	\S\S\ref{sect: results} and \ref{sect: discussion}. We summarise the main conclusions in \S\ref{sect: conclusions}.


\section{Observations}\label{sect: observations}
   	
	IRS~63 and Elias~29 were observed with the SMA\footnote{The Submillimeter Array is a joint project between the Smithsonian Astrophysical 
	Observatory and the Academia Sinica Institute of Astronomy and Astrophysics and is funded by the Smithsonian Institution and the Academia Sinica.} 
	\citep{ho:2004} on 15 and 17~May~2006, respectively. Weather conditions were good on 15~May, with zenith optical depths at 225~GHz 
	$\tau_{\rm 225}$ = 0.04--0.06 (as measured by a tipping radiometer operated by the Caltech Submillimeter Observatory). Conditions were slightly worse
	on 17~May, with $\tau_{\rm 225}$ starting at 0.16, falling to 0.1 during 
	the first half of the night. Physical baselines ranged from 11.6 to 62.0~metres, and the resulting projected baselines ranged from 
	5.9 to 63.8~k$\lambda$, yielding a resolution of about $4.0 \times 2.3\arcsec$ (natural weighting).
	The correlator was configured to observe the lines of HCO$^+$ $J$ = 3--2 and HCN $J$ = 3--2 together with the continuum emission at 268 and 278~GHz.
	The line data were taken with 512 channels of 0.2~MHz width each, resulting in a velocity resolution of 0.23~km s$^{-1}$.
	
	Calibration was done using the MIR package \citep{qi:2005}, and imaging was done using the MIRIAD package \citep{sault:1995}. The quasars
	\object{QSO B1622-297} and \object{QSO B1514-241} served as gain calibrators, and the absolute fluxes were calibrated on Uranus. The absolute flux 
	calibration is estimated to have an uncertainty of $\sim$20 \%. 
	The passbands were calibrated on \object{Callisto} and the quasar \object{QSO B1253-055} (\object{3C 279}).

\section{Results}\label{sect: results}
	
\begin{figure}
		\centering
		\includegraphics[width=\columnwidth]{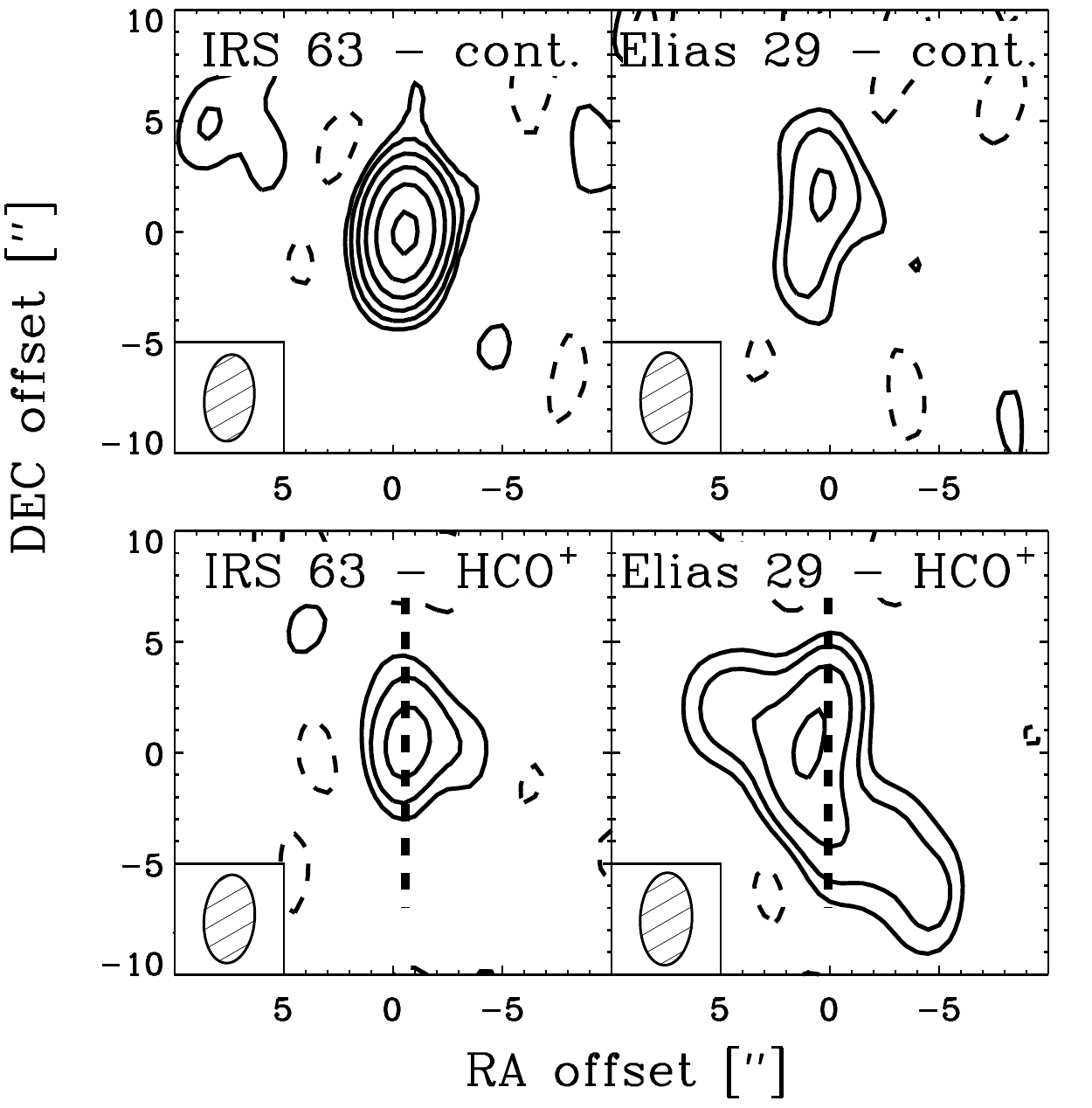}
		\caption[]{SMA images of the 1.1 mm continuum (top panels) and integrated HCO$^+$ 3--2 (bottom panels) emission. Contours 
		are drawn at 2, 4, 8, 16, 32, and 64 times the respective RMS noise levels, with the RMS being about 6~mJy~bm$^{-1}$ for 
		the continuum and about 65~mJy~bm$^{-1}$ for the line data. Negative contours are dashed. The positional offsets are with 
		respect to the coordinates of the observational phase centre. The dashed lines indicate how the position velocity diagrams 
		(Fig.~\ref{fig: PV diagrams}) are directed.}
		\label{fig: contour plots}
	\end{figure}

	The basic results of the observations are shown in Fig.~\ref{fig: contour plots} and summarised in Table~\ref{tab: results}. 
	
\subsection{Continuum data}\label{sect: results continuum}
	
	Both sources are clearly detected in the continuum at 1.1~mm. 
	Point-source and Gaussian fits were done in the ($u, v$) plane to determine the peak and integrated fluxes, respectively.
	A continuum peak flux of 459~mJy~bm$^{-1}$ and an integrated flux of 474~mJy were found for IRS~63, at a position which
	agrees with the 2MASS K$_{\rm S}$-band position within $0\farcs1$. IRS~63 was also observed with the SMA by \citet{andrews:2007} at higher resolution at
	1.3~mm. Their interferometer flux for IRS~63, to which they refer as \object{L1709~B}\footnote{IRS~63 and L1709~B are often referred to as the same
	source \citep[e.g.,][]{andre:1994}. However, L1709~B would appear to lie about 4~arcmin north of IRS~63 \citep[e.g.,][]{benson:1989}. The source that
	\citet{andrews:2007} refer to as L1709~B, is in fact IRS~63.}, is consistent with the value
	here and a mm slope $\alpha_{\rm mm} = 2.0 \pm 1.0$.
	The data of Elias~29 are best fitted by two sources. The brightest of the two agrees with the 2MASS K$_{\rm S}$-band position 
	of Elias~29 within $0\farcs2$, whereas the other source is offset by 3$\arcsec$. The second peak is attributed to an enhancement in 
	the ridge, from which Elias~29 likely formed, and is designated by ``Ridge'' in Table~\ref{tab: results}. Elias 29 and the ridge 
	component yield continuum peak fluxes of 47 and 31~mJy~bm$^{-1}$, and integrated fluxes of 72 and 28~mJy, respectively.
	
\begin{figure}
		\centering
		\includegraphics[width=\columnwidth]{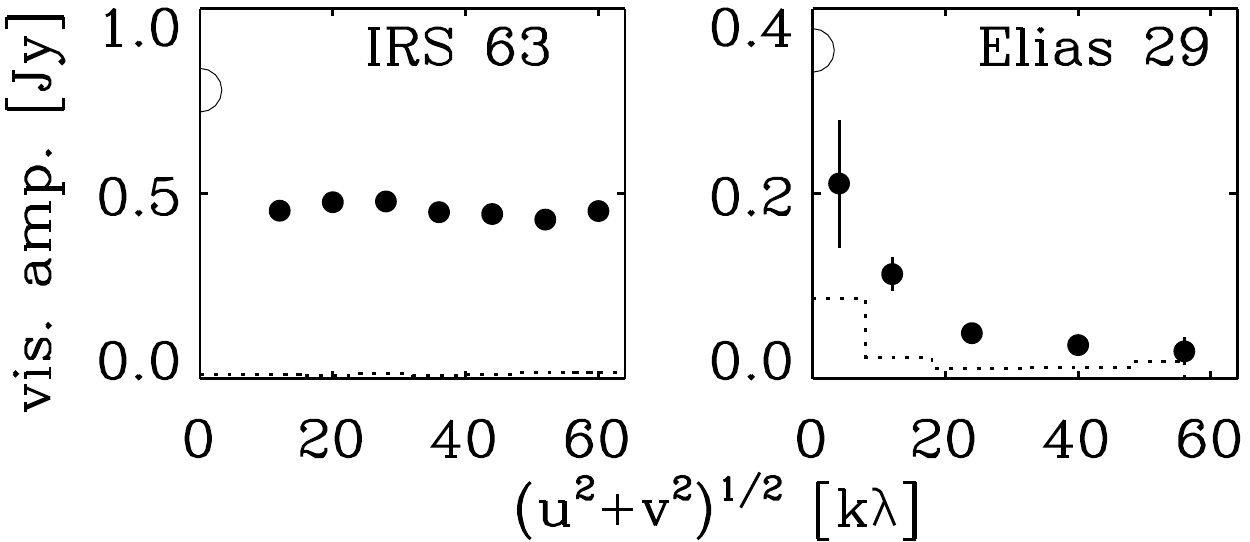}
		\caption[]{Continuum amplitude as a function of projected baseline for IRS~63 (left panel) and Elias~29 (right panel). 
		The data points give the amplitude per bin, where the data are binned in annuli according to ($u, v$) distance. The error 
		bars show the statistical $1\sigma$ errors, most often smaller than the data points, and the dotted lines give the expected 
		amplitude for zero signal, i.e., the anticipated
		amplitude in the absence of source emission. The half-open circles at zero ($u, v$) distance give the zero-spacing 1.1~mm flux, interpolated 
		between 850~$\mu$m and 1.25~mm single-dish fluxes \citep[][see text]{andre:1994, johnstone:2004, ridge:2006}.}
		\label{fig: uvamps}
	\end{figure}

	Plots of the visibility amplitudes as functions of projected baseline for
	IRS~63 and Elias~29 are shown in Fig.~\ref{fig: uvamps}. Elias~29 shows
	a steeply rising flux toward the shortest baselines, indicating that an envelope is
	present around this source. The emission toward IRS~63 does not change appreciably with
	baseline length, indicating that this source is unresolved up to
	baselines of 60~k$\lambda$, or down to physical scales of about 
	400~AU.
	
	
	Fig.~\ref{fig: uvamps} also shows the estimated single-dish flux at 1.1~mm. For this, the ``mapped'' fluxes from \citet{andre:1994} 
	at 1.25~mm were combined with newly determined 850~$\mu$m fluxes \citep[SCUBA maps from the COMPLETE survey,][]{johnstone:2004, 
	ridge:2006}. The SEDs are assumed to have a $F_\nu \propto \nu^\alpha$ dependence in this wavelength range, and are interpolated to 
	the 1.1~mm wavelength at which the SMA observations were conducted. Fluxes of 355 and 780 mJy are found for Elias~29 and IRS~63, 
	respectively, with an estimated uncertainty of 25\%. These fluxes are significantly larger than	the interferometer fluxes and 
	indicate that extended emission is present around the compact components picked up with the SMA. This extended emission may be due 
	to a circumstellar envelope, or to surrounding or intervening interstellar clouds, since the ``mapped'' regions of 
	\citet{andre:1994} are rather large, 40$\arcsec$ for Elias~29 and 60$\arcsec$ for IRS~63.

\subsection{Line data}\label{sect: results line}

\begin{figure}
		\centering
		\includegraphics[width=\columnwidth]{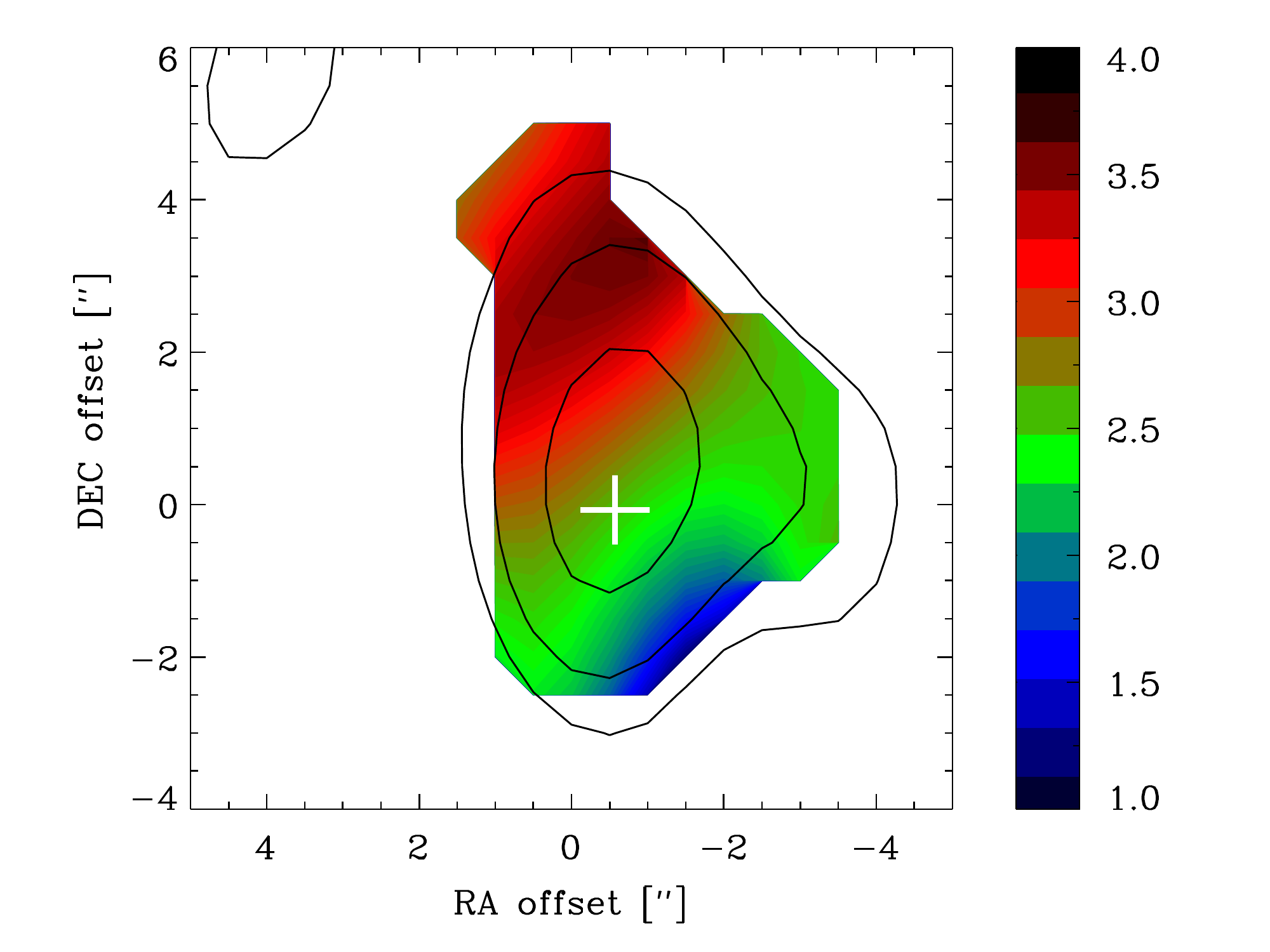}
		\caption[]{First-moment map of IRS~63. Contours are drawn at 2, 4, and 8 times the RMS of 67~mJy~bm$^{-1}$. The positional offsets are with
		respect to the coordinates of the observational phase centre, and the cross indicates the continuum position of IRS~63 from a point-source fit
		in the ($u, v$) plane, see Table~\ref{tab: results}. {\em A colour version of this figure can be obtained from the electronic version of the
		paper.}}
		\label{fig: moment map}
	\end{figure}

\begin{figure}
		\centering
		\includegraphics[width=\columnwidth]{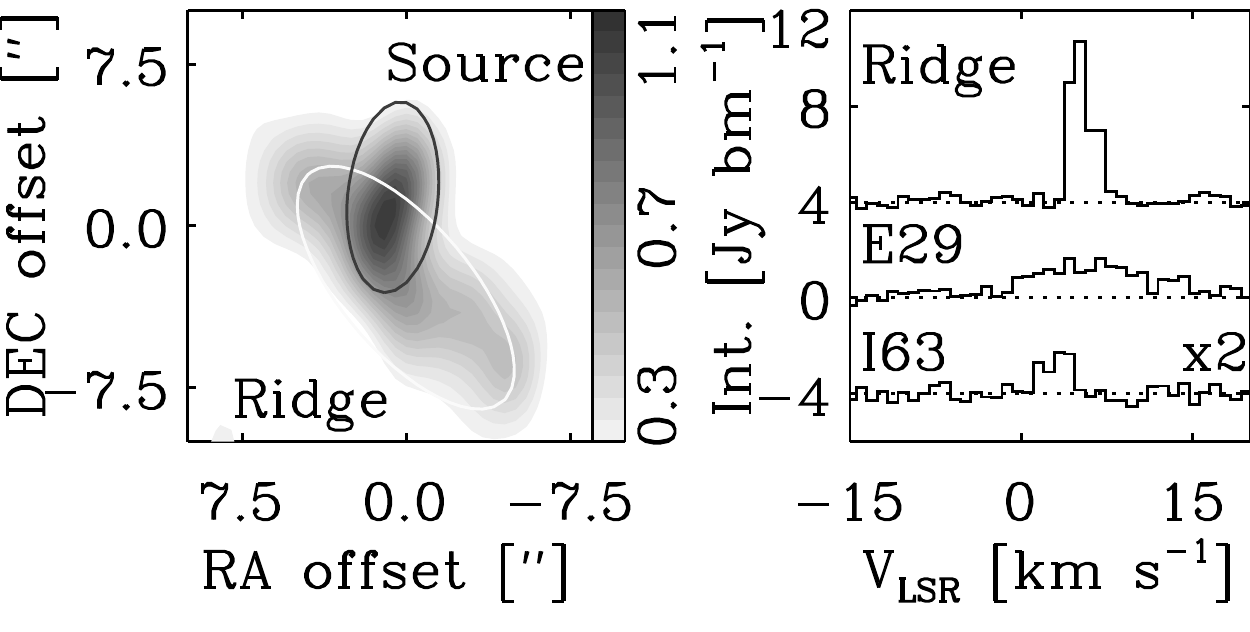}
		\caption[]{Left panel: HCO$^+$ $J$ = 3--2 integrated emission toward Elias~29. 
		The grayscale is linear from 0.2 to 1.2~Jy~bm$^{-1}$. The emission is fitted with two Gaussians, the one in the north-south direction 
		tracing the emission due to Elias~29, and the other, larger one tracing
		emission from a density enhancement in the ridge. Right panel: spectral lines, integrated over $2 \times 2\arcsec$ regions, centred
		on the continuum positions of Elias~29, of the ridge (shifted by +4~Jy~bm$^{-1}$), and of IRS~63 (shifted by -4~Jy~bm$^{-1}$),
		and binned to 0.9 km~s$^{-1}$.
		The HCO$^+$
		$J$ = 3--2 emission toward IRS~63 appears to be centred at V$_{\rm LSR} = 3.3$~km~s$^{-1}$.}
		\label{fig: line}
	\end{figure}

	HCO$^+$ $J$ = 3--2 is detected toward both sources, whereas HCN $J$ = 3--2 is not. Elias~29 appears to be about ten times as 
	strong in HCO$^+$ $J$ = 3--2 integrated emission as IRS~63 (see Table~\ref{tab: results}). 
	The emission toward IRS~63 is compact, coincident with the continuum peak, and shows a velocity gradient in the north-south 
	direction, see Fig.~\ref{fig: moment map}.
	The integrated emission in the direction of Elias~29 is more extended and can be
	fitted with two elliptical Gaussians, as is shown in the left panel of Fig.~\ref{fig: line}. The centre position of the 
	smaller Gaussian coincides with the continuum peak, as well as with 
	the infrared position, and is attributed
	to Elias~29 itself. The second Gaussian has a very elongated shape, offset from the infrared source, and is attributed to a density enhancement in 
	the nearby ridge.
	The right panel of Fig.~\ref{fig: line} shows the HCO$^+$ $J$ = 3--2 spectra toward the continuum positions of Elias~29 and IRS~63, binned to 0.9~km~s$^{-1}$.
	The HCO$^+$ $J$ = 3--2 emission toward IRS~63 appears to be centred at $V_{\rm LSR} = 3.3$~km~s$^{-1}$.
	
	\begin{table}
		\caption[]{Results of SMA observations at 1.1~mm.}
		\label{tab: results}
		\centering
		\begin{tabular}{lccc}
			\hline \hline
									& IRS~63		& Elias~29		& Ridge			\\
			\hline
			$F_\nu$ (P)$^\mathrm{a}$ [mJy~bm$^{-1}$]	& 459			& 47			& 31			\\
			RA$^\mathrm{a}$ [J2000]				& 16:31:35.65		& 16:27:09.42		& 16:27:09.48		\\
			Dec$^\mathrm{a}$ [J2000]			& -24:01:29.56		& -24:37:18.91		& -24:37:22.48		\\
			RMS$^\mathrm{b}$ [mJy~bm$^{-1}$]		& 6			& 5			& 5			\\
			\hline
			\multicolumn{4}{l}{Circular Gaussian fits to all data}									\\
			\hline
			$F_\nu$ (G)$^\mathrm{c}$ [mJy]			& 474			& 72			& 28			\\
			FWHM$^\mathrm{c}$ [arcsec]			& $0.55 \pm 0.08$	& $2.2 \pm 0.5$		& --			\\
			\hline
			\multicolumn{4}{l}{Circular Gaussian fits to data $\ge$16~k$\lambda$ (scales $< 13$~arcsec)}				\\
			\hline
			$F_\nu$ (G)$^\mathrm{c}$ [mJy]			& 480			& 57			& 27			\\
			FWHM$^\mathrm{c}$ [arcsec]			& $0.62 \pm 0.08$	& $1.7 \pm 0.7$		& --			\\
			\hline
			\multicolumn{4}{l}{Line data}												\\
			\hline
			HCO$^+$$^\mathrm{d}$ [K km s$^{-1}$]		& $3.1 \pm 0.3$		& $23.7 \pm 2.5$	& $23.5 \pm 2.5$	\\
			HCN$^\mathrm{d}$ [K km s$^{-1}$]		& $< 1.8$		& $< 3.8$		& $< 3.8$		\\
			\hline
		\end{tabular}
		\begin{list}{}{}
			\item[$^\mathrm{a}$] Point source fit in the ($u, v$) plane.
			\item[$^\mathrm{b}$] Calculated from the cleaned image.
			\item[$^\mathrm{c}$] Circular Gaussian fit in the ($u, v$) plane.
			\item[$^\mathrm{d}$] Integrated intensities are from a $4 \times 4\arcsec$ square around the continuum emission. 
			The synthesised beam is $\sim 4 \times 2.5\arcsec$. HCN was not detected; quoted values are 3$\sigma$ upper limits.
		\end{list}
	\end{table}

\section{Discussion and interpretations}\label{sect: discussion}

\subsection{Envelope and disk masses}\label{sect: envelope and disk masses}

	The basic data as presented in the previous section are used to determine the masses of the disks and the envelopes in the systems, both empirically
	(Sect.~\ref{sect: empirical results}) and through detailed radiative-transer modelling (Sect.~\ref{sect: modelling}). The differences between the two
	methods and the over-all conclusions are discussed in Sect.~\ref{sect: discussion of envelope and disk masses}.

\subsubsection{Empirical results}\label{sect: empirical results}
	
	A first-order estimate
	of the disk and envelope masses can be obtained by assuming that the continuum flux of the envelope is resolved out
	by the interferometer on the longest baselines (Fig.~\ref{fig: uvamps}).
	Hence, the emission on the longest baselines is solely due to the disk, whereas the single-dish flux comes from the disk and envelope combined.
	Assuming that the disk and the envelope are isothermal and the dust emission is optically thin, the disk mass is given by
	\begin{equation}\label{eq: disk mass}
		M_{\rm disk} = \frac{F_\nu \Psi D^2}{\kappa_\nu B_\nu(T_{\rm dust})},
	\end{equation}
	where $F_\nu$ is the flux at frequency $\nu$ on the longest baselines, $\Psi$ is the gas-to-dust ratio, $\kappa_\nu$ is the dust opacity at $\nu$, 
	and $B_\nu(T_{\rm dust})$ is the emission from a black body at $T_{\rm dust}$. 
	Using fiducial values for all these parameters -- $\Psi = 100$, $T_{\rm dust} = 30$~K, and $\kappa_\nu = 1.18$~cm$^2$~g$^{-1}$ \citep[``OH5''
	coagulated dust with icy mantles, found in the fifth column of Table~1 of][interpolated to the observed frequency 
	$\nu = 273$~GHz]{ossenkopf:1994} -- disk masses of $0.055$~M$_\odot$ and $0.007$~M$_\odot$ are found for IRS~63 and Elias~29, respectively.
	Using the opacity law as quoted by \citet{beckwith:1990}, $\kappa_\nu = 10 
	\left( \nu/10^{12} {\rm Hz} \right) ^\beta$, and taking $\beta = 1$, a dust opacity at 273~GHz of $\kappa_\nu = 2.73$~cm$^2$~g$^{-1}$ 
	is found, and the disk masses decrease to 
	$0.024$~M$_\odot$ for IRS~63 and $0.003$~M$_\odot$ for Elias~29.\footnote{The opacity law quoted by 
	\citet{beckwith:1990} is $\kappa_\nu = 0.1 \left( \nu/10^{12} {\rm Hz} \right) ^\beta$ for the gas and dust 
	combined, implicitly assuming a gas-to-dust ratio $\Psi = 100$, consistent with our work.}
	Note that this method of estimating the disk mass is only valid when the disk is unresolved by the observations.
	If the disk were clearly resolved, the flux on the longest baselines would not include the emission from the outer disk, but only that from 
	smaller radii, unresolved by those baselines. Due to its closeness, the enhancement in the ridge near Elias~29 may
	contribute to its flux, even on the longest baselines of $\ge 50$~k$\lambda$ or scales $< 4\arcsec$. 
	However, in the estimate of the flux on the longest baselines (Table~\ref{tab: results}), the ridge component was fitted simultaneously with Elias~29,
	and hence its contribution to the derived disk mass is expected to be minimal.
	
	In a similar fashion the envelope masses in the systems can be found. First, the contributions from the envelopes to the 1.1~mm single-dish
	fluxes are determined by subtracting the fluxes at baselines $\ge 16$~k$\lambda$ (see Table~\ref{tab: results}) from the single-dish fluxes as found
	in Sect.~\ref{sect: results}. The envelope masses are then found by using Eq.~(\ref{eq: disk mass}), with the temperatures of the envelopes taken 
	to be 20~K. For an optically thin envelope around stars with luminosities of 0.79 and 13.6~L$_\odot$ applicable to IRS~63 and Elias~29, a 
	temperature of 17--29~K
	is expected at a radius of 940~AU corresponding to the size of the JCMT single-dish beam \citep[from, e.g., Eq.~(2) of ][taking the opacity index
	$\beta = 1$]{chandler:2000}. This yields an envelope mass $M_{\rm env} = 0.058$~M$_\odot$ for IRS~63 within a 30$\arcsec$ radius, where the 
	contribution from the envelope to the total 1.1~mm flux is 38\%. Coincidentally, also for Elias~29 an envelope 
	mass $M_{\rm env} = 0.058$~M$_\odot$, within a 20$\arcsec$ radius, is found. However, in this system
	the contribution from the envelope to the total 1.1~mm flux is 84\%. As for the disk mass, also the envelope mass found for Elias~29 should be taken
	as an upper limit because of the closeness of the ridge enhancement. Note, however, that it is unlikely that the bulk of the continuum emission 
	is due to this enhancement, because of the coincidence of the continuum peak with the infrared source and also with the single-dish 
	submillimetre positions. Using the opacities of \citet{beckwith:1990} will lower all masses by a factor of about 2.
	
\subsubsection{Modelling}\label{sect: modelling}	

	To provide more physical grounding to the values for $M_{\rm disk}$ and $M_{\rm env}$ estimated in the previous section, a radiative transfer code was used to fit
	simultaneously the millimetre interferometry and the spectral energy distributions. IRS~63 and Elias~29 were modelled with a new version of the programme RADMC
	\citep[e.g.,][]{dullemond:2004}. RADMC is an axisymmetric Monte-Carlo code for dust continuum radiative transfer in circumstellar disks and envelopes in which the 
	stellar	photons are traced in three dimensions. The code is based on the method of \citet{bjorkman:2001}.
	
	The density structure of the disk is given by \citep[see][]{crapsi:2007}
	\begin{equation}\label{eq: disk structure}
		\rho_{\rm disk}(r, \theta) = \frac{\Sigma_0(r/R_0)^{-1}}{\sqrt(2\pi)H(r)}\exp \left( - \frac{1}{2}\left[\frac{r \cos(\theta)}{H(r)}\right]^2\right),
	\end{equation}
	where $r$ is the radial distance from the central star and $\theta$ is the angle from the axis of symmetry. The inner radius was fixed to 0.1~AU for IRS~63, and to
	0.25~AU for Elias~29, to account for the higher luminosity of the star. At these radii, the temperature is of the order of the dust sublimation temperature.
	Note that the exact sublimation temperature, and hence the disk inner radius, depends on the exact dust species and local density. The vertical scale height of the disk 
	is given by $H(r) = r \cdot H_0/R_0 \cdot (r/R_0)^{0.17}$, where $R_0$ was left free to vary, and $H_0$/$R_0$ was arbitrarily fixed to 0.17.
	\citet{chiang:1997} find $H(r)/r \propto r^{2/7}$. However, that estimate is based on grey opacities and constant surface density, and somewhat smaller
	flaring is usually more realistic. Hence, we chose $H(r) \propto r \cdot r^{0.17}$.
	$\Sigma_0$ was scaled to the disk mass, which was also left as a free parameter.
	The envelope density follows the equation for a rotating, infalling model \citep[][]{ulrich:1976, crapsi:2007}
	\begin{eqnarray}\label{eq: envelope structure}
		\mbox{} \qquad \rho_{\rm envelope}(r, \theta)=\rho_0 \left( \frac{r}{R_{\rm rot}} \right)^{-1.5} \left( 1+\frac{\cos{\theta}}{\cos{\theta_0}} \right)^{0.5} \times \nonumber\\
		\mbox{} \qquad \times \left( \frac{\cos{\theta}}{2 \cos{\theta_0}} + \frac{R_{\rm rot}}{r} \cos^2 \theta_0 \right)^{-1},
	\end{eqnarray}
	where $\theta_0$ is the solution of the parabolic motion of an infalling particle, $R_{\rm rot}$ is the centrifugal radius of the envelope, and 
	$\rho_0$ is the density in the equatorial plane at $R_{\rm rot}$. $\rho_0$ was scaled to 
	accomodate the total envelope mass, and $R_{\rm rot}$, which can have a significant influence on the amplitude as a function of baseline length, 
	was left free to vary. The outer radius of the envelope was fixed to 10,000~AU, where the temperature is similar to that of the ambient interstellar cloud.
	The parameters, as summarised in Table~\ref{tab: model params}, give the best-fit models by eye; a full $\chi^2$ 
	minimization is not warranted by the relatively low resolution and S/N of the data. 
	Our models do not have outflow cavities in the envelopes. Note that the actual physical structure 
	of the envelope (centrifugal radius, outflow cavities, etc.) may considerably affect the mid-IR SED and thus the derived inclination from the 
	models \citep[e.g.,][]{whitney:2003, jorgensen:2005}.
	\begin{table}
		\caption[]{Model parameters for IRS~63 and Elias~29.}
		\label{tab: model params}
		\centering
		\begin{tabular}{lcc}
			\hline \hline
										& IRS~63			& Elias~29			\\
			\hline
			\multicolumn{3}{l}{Modelling parameters and results}									\\
			\hline
			Stellar luminosity ({\it fixed})			& 0.79~L$_\odot$		& 13.6~L$_\odot$		\\
			Disk inner radius ({\it fixed})				& 0.1~AU			& 0.25~AU			\\
			Disk outer radius, $R_0$				& 100~AU			& 200~AU			\\
			Disk height at outer radius, $H_0$/$R_0$ ({\it fixed})	& 0.17				& 0.17				\\
			Envelope outer radius ({\it fixed})			& 10,000~AU			& 10,000~AU			\\
			Centrifugal radius, $R_{\rm rot}$			& 100~AU			& 300~AU			\\
			Inclination						& 30$^\circ$			& 30$^\circ$			\\
			Disk mass, $M_{\rm disk}$				& 0.13~M$_\odot$		& 0.004~M$_\odot$		\\
			Envelope mass, $M_{\rm env}^\mathrm{a}$			& 0.022~M$_\odot$		& 0.025~M$_\odot$		\\
			\hline
			\multicolumn{3}{l}{Empirical results}											\\
			\hline
			$M_{\rm env}^\mathrm{a}$ [M$_\odot$]			& $0.058$			& $\leq 0.058$			\\
			$M_{\rm disk}$ [M$_\odot$]				& $0.055$			& $\leq 0.007$			\\
			$M_{\rm star}^\mathrm{b}$ [M$_\odot$]			& $0.37 \pm 0.13$		& $2.5 \pm 0.6$			\\
			\hline
		\end{tabular}
		\begin{list}{}{}
			\item[$^\mathrm{a}$] The envelope mass is defined as the mass, not contained in the disk, within a single-dish radius, i.e., 30$\arcsec$/3750~AU
						for IRS~63, and 20$\arcsec$/2500~AU for Elias~29.
			\item[$^\mathrm{b}$] Assuming an inclination $i = 30^\circ$ ($\S~4.2$ for discussion).
		\end{list}
	\end{table}
	
	Two different dust species are used in these models, ``hot'' 
	dust where the temperature is higher than 90~K, and ``cold'' dust where the temperature is below 90~K.
	The opacities used are ``OH2''
	\citep[coagulated dust without ice mantles,][]{ossenkopf:1994} for the hot dust, and ``OH5'' (coagulated dust with thin ice mantles) for the cold 
	dust. The central stars were represented by sources with luminosities of 0.79~$L_\odot$ for IRS~63 and 13.6~L$_\odot$ for Elias~29, which were 
	taken to be equal to the bolometric luminosities of the sources, and the temperature structures in the disks and envelopes were then calculated by 
	propagating the stellar photons through the systems.

\begin{figure}
		\centering
		\includegraphics[width=\columnwidth]{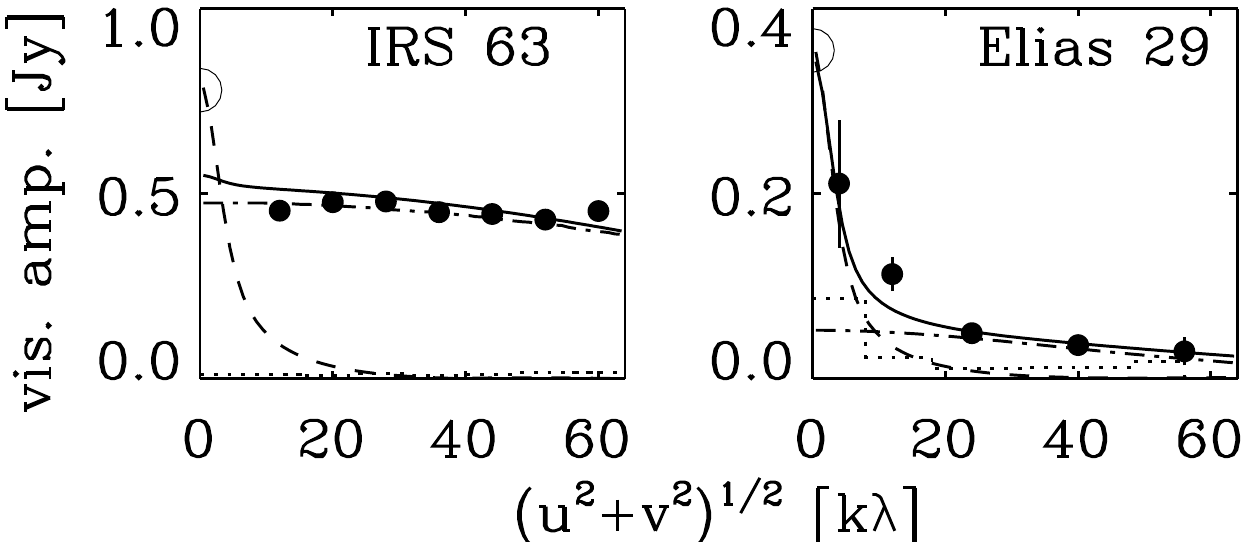}
		\caption[]{Continuum amplitude as a function of projected baseline for IRS~63 (left panel) and Elias~29 (right panel), with models
		overplotted. The dashed lines show models, consisting of an envelope only. They fit the single-dish (sub)mm data well, but fall off too
		fast to longer projected baselines. The dash-dotted lines are from disk-only models. They fit the interferometric data well, but do not
		have enough flux on the shortest baselines. The solid lines are for the models that provide a good fit to the SED and to the spatial
		information provided by the interferometric observations. The dotted lines indicate the expected amplitude for zero signal.}
		\label{fig: models}
	\end{figure}

\begin{figure}
		\centering
		\includegraphics[width=\columnwidth]{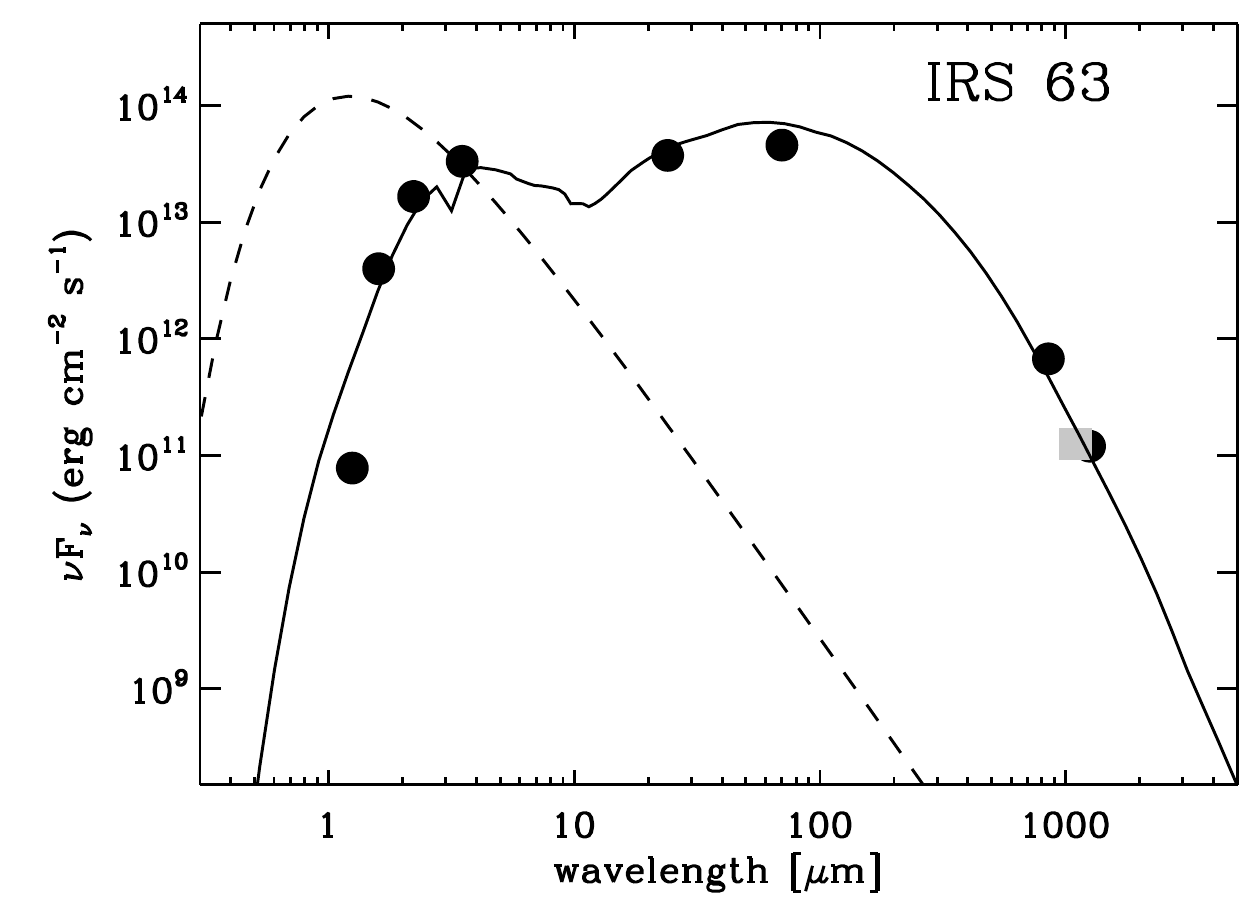}
		\caption[]{Spectral energy distribution of IRS~63 using the model parameters of Table~2 extincted by A$_{\rm V}$=7 mag
		(solid line). The dashed line shows the model star, which has a luminosity of 0.79~L$_\odot$, and is represented by a 
		black-body with a temperature of 3~000~K.
		Overplotted are $JHKL$ photometry \citep{haisch:2002}, MIPS 24 and 70~$\mu$m fluxes 
		(c2d\protect\footnotemark), 
		JCMT 850~$\mu$m 60$\arcsec$ integrated flux (COMPLETE), SMA 1.10~mm peak flux (this work; square), and IRAM 1.25~mm 60$\arcsec$ 
		integrated flux \citep{andre:1994}.}
		\label{fig: irs63 sed}
	\end{figure}

	The results of the modelling are shown in Fig.~\ref{fig: models}. 
	A model with only a 0.055~M$_\odot$ disk reproduces the amplitude as a function of baseline quite well for IRS~63, but fails to 
	produce enough single-dish flux. A 0.35~M$_\odot$ envelope-only model, on the other hand, produces enough single-dish flux in the (sub)mm regime, 
	but falls off
	too quickly toward longer baselines. 
	To allow for a comparison with the empirical model (Sect.~\ref{sect: empirical results}) the envelope mass was defined as the mass, present within the single-dish
	aperture (i.e., a 30$\arcsec$ radius for IRS~63 and a 20$\arcsec$ radius for Elias~29), and not contained in the disk.
	A model with both a disk and an envelope is necessary to explain the observations of IRS~63, and the best model 
	consists of a disk of 0.13~M$_\odot$ and an envelope of 0.022~M$_\odot$. This model also fits the observed IRS~63 SED well, if a foreground 
	extinction of A$_{\rm V} = 7$~mag is assumed, see Fig.~\ref{fig: irs63 sed}. 
	The region in which IRS~63 is located has a very high A$_{\rm V}$ of about 24 mag as found in extinction maps produced by the Spitzer 
	``Cores to Disks'' Legacy program \citep[][]{evans:2003,evans:2007}. These maps measure the large scale ($\sim 5\arcmin$) cloud extinction on 
	the basis of background stars, and therefore includes an extra contribution besides envelope extinction.	

	The amplitude as a function of baseline for Elias~29 is best explained by a 0.004~M$_\odot$ disk and a 0.025~M$_\odot$ envelope. For this source, 
	no further attempt to model the SED was made, because the close presence of the ridge enhancement makes a model that is axisymmetric on scales 
	larger than several 100~AU unapplicable.
	\addtocounter{footnote}{-1}
	\addtocounter{footnote}{-1}
	\stepcounter{footnote}\footnotetext{Delivery of data from the c2d Legacy Project: IRAC and MIPS \citep[Pasadena, SSC,][]{evans:2007}.}

\subsubsection{Discussion}\label{sect: discussion of envelope and disk masses}

	Even though the disk and envelope masses derived from the empirical and detailed model results differ by up to a factor of a few, we are confident
	that they are accurate within this range. For example, a significantly larger disk in the IRS~63 model would show a very steep fall-off in 
	the amplitude as a function of baseline length, which is not observed (Fig.~\ref{fig: models}). Likewise, an envelope significantly more massive than 0.022~M$_\odot$ found 
	for IRS~63 would obscure the central star so much that the near- and mid-IR flux would be severly underestimated (Fig.~\ref{fig: irs63 sed}). In the further discussion, we 
	will use the disk and envelope masses as found by the detailed modelling, but note that the overall conclusions do not change if the empirical results are used.

	The envelope masses of IRS~63 and Elias~29 are in the same range as those presented by, e.g.,
	\citet{hogerheijde:1997a} for a sample of embedded Class I sources in the Taurus-Auriga star-forming region. However, the envelope masses 
	of the deeply embedded Class 0 objects are found to be considerably higher, $\gtrsim 1$~M$_\odot$ 
	\citep[][]{jorgensen:2002, shirley:2002, young:2003, hatchell:2007}.
	While some fraction of sources may simply originate from lower mass cores, it is clear that all Class 0 objects must pass through a stage with
	$M_{\rm env} \approx 0.1$~M$_\odot$ on their way to the pre-main sequence stage. A larger sample is needed to address the question
	whether the mass that was initially present in the envelope will first accumulate in the disk, or whether it will pass through the disk and onto the
	star directly, leaving the disk in a steady state through the entire Class I stage.
	
	An interesting quantity for these YSOs is the ratio $M_{\rm env}$/$M_{\rm disk}$, as this may be a direct measure for their evolutionary stage.
	The values for the two objects under consideration here -- 0.2 for IRS~63 and 6 for Elias~29 --  
	are quite different, although the value for Elias~29 is rather uncertain
	due to the contribution of the ridge enhancement to the continuum emission. 
	From these simple arguments, Elias~29 is the less evolved, rather like the deeply embedded Class~0 sources in the PROSAC sample \citep{jorgensen:2007}, which show 
	$M_{\rm env}$/$M_{\rm disk}\gtrsim 10$. IRS~63, on the other hand, is well on its way toward the optically visible Class II or T Tauri stage, 
	which have $M_{\rm env} \approx 0$.
	This confirms the notion that the Lada classification is primarily a phenomenological one, not necessarily representing the actual physical stage the YSO is in
	\citep[see, e.g.,][]{crapsi:2007}.

\subsection{Keplerian rotation and stellar masses}\label{sect: stellar masses}

\begin{figure}
		\centering
		\includegraphics[width=\columnwidth]{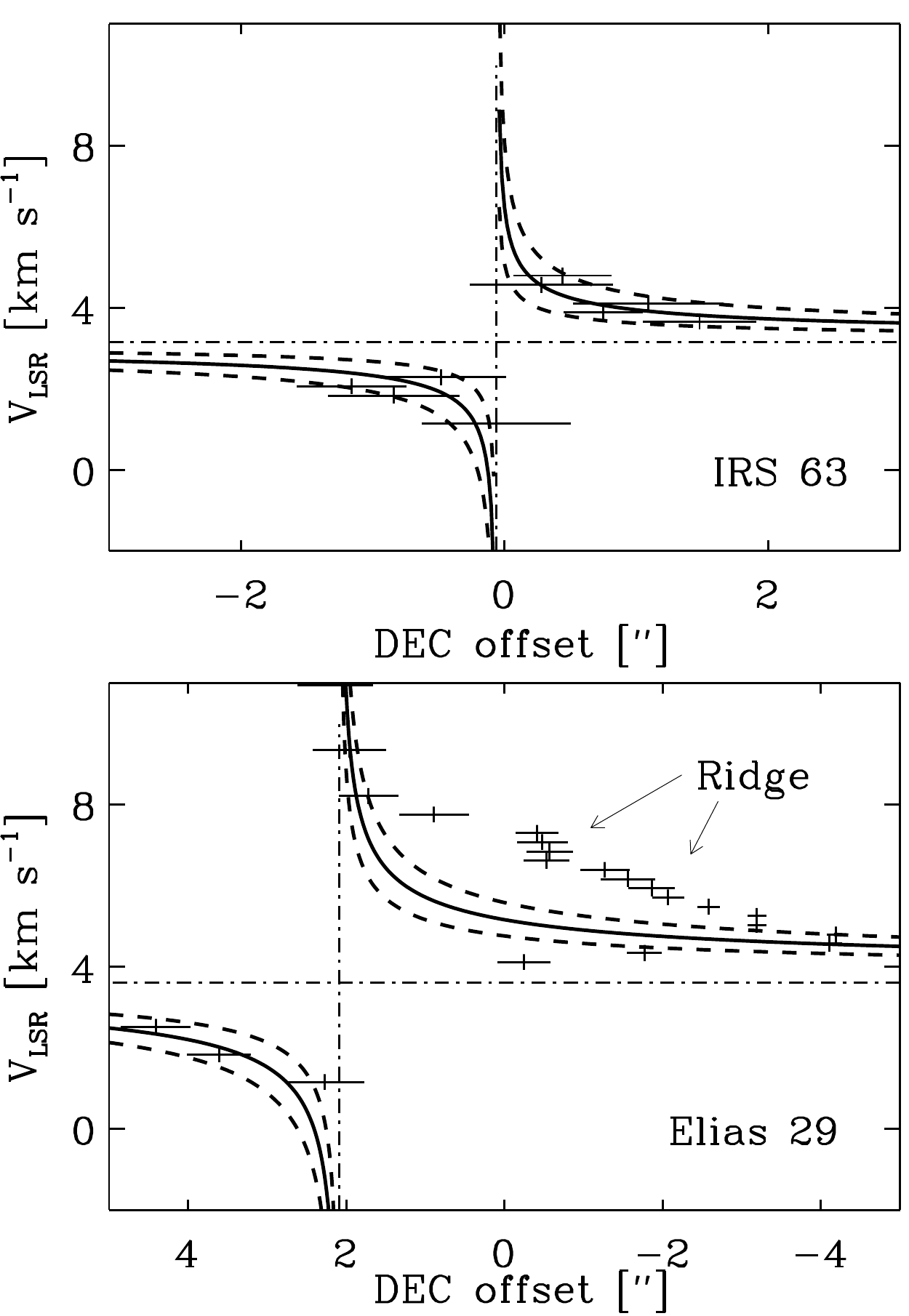}
		\caption[]{Velocity of the HCO$^+$ $J$ = 3--2 line as a function of the declination offset from the phase centre for
		IRS~63 ({\it upper panel}) and Elias~29 ({\it lower panel}). The solid lines show the results from $\chi^2$ fits to the data, yielding central masses of
		0.37~M$_\odot$ (IRS~63) and 2.5~M$_\odot$ (Elias~29) for inclinations of 30$^\circ$; the dashed lines show formal 99\% confidence intervals. 
		Points attributed to the emission from the dense ridge are indicated and are not included in the fit.
		The declination offsets are different for the two panels.}
		\label{fig: PV diagrams}
	\end{figure}

	Both IRS~63 and Elias~29 show a velocity gradient in the HCO$^+$ emission, which is interpreted as the rotation of a circumstellar disk.
	The mass of the central object can be estimated from position-velocity diagrams along
	the major axis of the disk, as indicated in Fig.~\ref{fig: contour plots}. 
	An elliptical Gaussian was fitted to each channel in the HCO$^+$ data in the ($u, v$) plane, using the MIRIAD task {\em uvfit}. This yielded a best-fit 
	position per channel with corresponding uncertainties in right ascension and declination, and consequently a
	declination offset from the phase centre in the physical plane. Note that the fitting was carried out in the ($u, v$) plane, so as not to be hindered by
	artefacts that may arise in the deconvolution process. Position-velocity diagrams were created by plotting the declination offset as a function of velocity
	channel, where only coordinates with an uncertainty in the declination of less than $0\farcs5$ were taken into account. Subsequently, the signatures of Keplerian
	rotation around a point source were fitted to the position-velocity diagrams
	(Fig.~\ref{fig: PV diagrams}). In the fitting process, the values for $M_{\rm star}$ (the mass of the central object), the central velocity of the
	system with respect to the local rest frame, and the declination offset from the source to the phase centre were varied, and a global minimum $\chi^2$
	was determined.  
	
	Assuming a Keplerian disk seen edge-on ($i = 90^\circ$), a central mass of $M_{\rm star} = 0.09 \pm 0.03$~M$_\odot$ is found for IRS~63.
	The exact inclination of IRS~63 is hard to determine due to the high extinction in the region. However, an inclination larger than $\sim 45^\circ$ can be ruled out
	because of the source's brightness at 3--5~$\mu$m \citep{pontoppidan:2003}.
	Assuming an inclination of 30$^\circ$, and taking 
	into account the $\sin^2 i$ dependence, the mass of the
	central object increases to $M_{\rm star} = 0.37 \pm 0.13$~M$_\odot$.
	For Elias~29, the points attributed to the emission from the dense ridge (Fig.~\ref{fig: PV diagrams}) were disregarded in the fitting 
	procedure. 
	A lower limit to the central mass of $M_{\rm star} = 0.62 \pm 0.14$~M$_\odot$ is found, under the assumption $i = 90^\circ$.
	The flatness of the SED limits the inclination of this source to $< 60^\circ$ \citep[][]{boogert:2002}. A fully face-on orientation,
	on the other hand, is unlikely, given the presence of low surface brightness scattered K-band light \citep{zinnecker:1988}. An inclination 
	$i = 30^\circ$ yields a mass for the central source of $2.5 \pm 0.6$~M$_\odot$. With such a high mass, the central star is likely to emerge as a
	Herbig~Ae/Be star, once the surrounding envelope is dissipated.

	It is interesting to compare the inferred masses to those inferred for optically visible pre-main sequence stars. With a stellar mass of 0.37~M$_\odot$ and a
	stellar luminosity of 0.79~L$_\odot$, IRS~63 would be found to have an age of about $5 \times 10^5$~yr, when compared to evolutionary tracks of
	classical and weak-lined T~Tauri stars in the
	$\rho$~Ophiuchi cloud \citep{wilking:2005,dantona:1997a}. A similar age would be found for Elias~29 \citep[e.g.,][]{palla:1993}. 

\section{Conclusions}\label{sect: conclusions}
	We used the SMA to observe the Class I YSOs Elias~29 and IRS~63 at 1.1~mm. The main results are as follows.
	\begin{itemize}
		\item Both sources are detected in the continuum and in the HCO$^+$ $J$ = 3--2 line, but not in the HCN $J$ = 3--2 line. The 
		HCO$^+$ emission toward IRS~63 is compact and associated with the YSO, whereas a significant part of the molecular emission in 
		the direction of Elias~29 is due to an enhancement in the ridge from which Elias~29 formed.
		\item Assuming that the continuum emission toward the sources is optically thin, and that any circumstellar envelope is resolved out by
		the interferometer on the longest baselines, we find empirical disk masses of 0.055 for IRS~63 and $\le 0.007$~M$_\odot$ for Elias~29.
		By comparing our interferometer data with single-dish observations, envelope masses of 0.058~M$_\odot$ are derived for
		both sources, within 30$\arcsec$ radius for IRS~63 and 20$\arcsec$ radius for Elias~29, where the value for Elias~29 should be treated as 
		an upper limit because of the presence of the ridge enhancement.
		\item Using a radiative-transfer code to create SEDs and amplitude vs. ($u, v$)-distance plots for different combinations of disk and
		envelope masses, we find disk masses of 0.13~M$_\odot$ for IRS~63 and 0.004~M$_\odot$ for Elias~29, and envelope masses
		of 0.022~M$_\odot$ and 0.025~M$_\odot$.
		\item The modelling results yield rather different ratios of M$_{\rm env}$/M$_{\rm disk}$: 0.2 for IRS~63 and 6 for Elias~29,
		with uncertainties of a factor of a few.
		From this it is concluded that IRS~63 is well on its way to become an optically visible Class~II source, rather than an embedded Class~I source. Elias~29, on
		the other hand, is more like the deeply embedded Class~0 sources.
		\item Velocity gradients in the HCO$^+$ $J$ = 3--2 emission are interpreted as signs of Keplerian rotation, and indicate central masses
		of $2.5 \pm 0.6$~M$_\odot$ for Elias~29 and $0.37 \pm 0.13$~M$_\odot$ for IRS~63, for inclinations of 30$^\circ$. These masses correspond to ages of
		a few $10^5$~yr.
	\end{itemize}

	With a larger set of sources treated in a similar manner as is presented for IRS~63 and Elias~29 here, it will be possible to 
	constrain models for the evolution of the envelope, disk, and stellar masses through this important stage of YSOs.

\begin{acknowledgements}
	We would like to thank the anonymous referee for a very constructive report, which greatly improved the paper.
	Demerese Salter and Doug Johnstone are thanked for providing us with data from JCMT SCUBA observations. The discussions with the members of the
	Leiden ``AstroChem'' group were very useful.
	We are grateful to Kees Dullemond for 
	providing us with the most recent version of the RADMC package. Partial support for this work was provided by a Netherlands Research School For 
	Astronomy network 2 grant, and by a Netherlands Organisation for Scientific Research Spinoza grant.
\end{acknowledgements}

\bibliographystyle{aa}
\bibliography{references}

\begin{thebibliography}{44}
\expandafter\ifx\csname natexlab\endcsname\relax\def\natexlab#1{#1}\fi

\bibitem[{{Andr\'{e}} \& {Montmerle}(1994)}]{andre:1994}
{Andr\'{e}}, P. \& {Montmerle}, T. 1994, \apj, 420, 837

\bibitem[{{Andr\'{e}} {et~al.}(1993){Andr\'{e}}, {Ward-Thompson}, \&
  {Barsony}}]{andre:1993}
{Andr\'{e}}, P., {Ward-Thompson}, D., \& {Barsony}, M. 1993, \apj, 406, 122

\bibitem[{{Andrews} \& {Williams}(2007)}]{andrews:2007}
{Andrews}, S.~M. \& {Williams}, J.~P. 2007, \apj, 659, 705

\bibitem[{{Beckwith} {et~al.}(1990){Beckwith}, {Sargent}, {Chini}, \&
  {Guesten}}]{beckwith:1990}
{Beckwith}, S.~V.~W., {Sargent}, A.~I., {Chini}, R.~S., \& {Guesten}, R. 1990,
  \aj, 99, 924

\bibitem[{{Benson} \& {Myers}(1989)}]{benson:1989}
{Benson}, P.~J. \& {Myers}, P.~C. 1989, \apjs, 71, 89

\bibitem[{{Bjorkman} \& {Wood}(2001)}]{bjorkman:2001}
{Bjorkman}, J.~E. \& {Wood}, K. 2001, \apj, 554, 615

\bibitem[{{Boogert} {et~al.}(2002){Boogert}, {Hogerheijde}, {Ceccarelli},
  {Tielens}, {van Dishoeck}, {Blake}, {Latter}, \& {Motte}}]{boogert:2002}
{Boogert}, A.~C.~A., {Hogerheijde}, M.~R., {Ceccarelli}, C., {et~al.} 2002,
  \apj, 570, 708

\bibitem[{{Chandler} \& {Richer}(2000)}]{chandler:2000}
{Chandler}, C.~J. \& {Richer}, J.~S. 2000, \apj, 530, 851

\bibitem[{{Chiang} \& {Goldreich}(1997)}]{chiang:1997}
{Chiang}, E.~I. \& {Goldreich}, P. 1997, \apj, 490, 368

\bibitem[{{Crapsi} {et~al.}(2007){Crapsi}, {van Dishoeck}, {Hogerheijde},
  {Pontoppidan}, \& {Dullemond}}]{crapsi:2007}
{Crapsi}, A., {van Dishoeck}, E.~F., {Hogerheijde}, M.~R., {Pontoppidan},
  K.~M., \& {Dullemond}, C.~P. 2007, A\&A in press.

\bibitem[{{D'Antona} \& {Mazzitelli}(1997)}]{dantona:1997a}
{D'Antona}, F. \& {Mazzitelli}, I. 1997, Memorie della Societa Astronomica
  Italiana, 68, 807

\bibitem[{{de Geus} {et~al.}(1989){de Geus}, {de Zeeuw}, \&
  {Lub}}]{degeus:1989}
{de Geus}, E.~J., {de Zeeuw}, P.~T., \& {Lub}, J. 1989, \aap, 216, 44

\bibitem[{{Dullemond} \& {Dominik}(2004)}]{dullemond:2004}
{Dullemond}, C.~P. \& {Dominik}, C. 2004, \aap, 417, 159

\bibitem[{{Evans} {et~al.}(2003){Evans}, {Allen}, {Blake}, {Boogert}, {Bourke},
  {Harvey}, {Kessler}, {Koerner}, {Lee}, {Mundy}, {Myers}, {Padgett},
  {Pontoppidan}, {Sargent}, {Stapelfeldt}, {van Dishoeck}, {Young}, \&
  {Young}}]{evans:2003}
{Evans}, N.~J., {Allen}, L.~E., {Blake}, G.~A., {et~al.} 2003, \pasp, 115, 965

\bibitem[{{Evans} {et~al.}(2007){Evans}, {Harvey}, {Dunham}, {Mundy}, {Lai},
  {Chapman}, {Huard}, {Bourke}, \& {Koerner}}]{evans:2007}
{Evans}, II, N.~J., {Harvey}, P.~M., {Dunham}, M.~M., {et~al.} 2007, Delivery
  of Data from the c2d Legacy Project: IRAC and MIPS (Pasadena, SSC)

\bibitem[{{Flaherty} {et~al.}(2007){Flaherty}, {Pipher}, {Megeath}, {Winston},
  {Gutermuth}, {Muzerolle}, {Allen}, \& {Fazio}}]{flaherty:2007}
{Flaherty}, K.~M., {Pipher}, J.~L., {Megeath}, S.~T., {et~al.} 2007, \apj, 663,
  1069

\bibitem[{{Greene} {et~al.}(1994){Greene}, {Wilking}, {Andre}, {Young}, \&
  {Lada}}]{greene:1994}
{Greene}, T.~P., {Wilking}, B.~A., {Andre}, P., {Young}, E.~T., \& {Lada},
  C.~J. 1994, \apj, 434, 614

\bibitem[{{Haisch} {et~al.}(2002){Haisch}, {Barsony}, {Greene}, \&
  {Ressler}}]{haisch:2002}
{Haisch}, Jr., K.~E., {Barsony}, M., {Greene}, T.~P., \& {Ressler}, M.~E. 2002,
  \aj, 124, 2841

\bibitem[{{Harvey} {et~al.}(2003){Harvey}, {Wilner}, {Myers}, \&
  {Tafalla}}]{harvey:2003}
{Harvey}, D.~W.~A., {Wilner}, D.~J., {Myers}, P.~C., \& {Tafalla}, M. 2003,
  \apj, 596, 383

\bibitem[{{Hatchell} {et~al.}(2007){Hatchell}, {Fuller}, {Richer}, {Harries},
  \& {Ladd}}]{hatchell:2007}
{Hatchell}, J., {Fuller}, G.~A., {Richer}, J.~S., {Harries}, T.~J., \& {Ladd},
  E.~F. 2007, \aap, 468, 1009

\bibitem[{{Ho} {et~al.}(2004){Ho}, {Moran}, \& {Lo}}]{ho:2004}
{Ho}, P.~T.~P., {Moran}, J.~M., \& {Lo}, K.~Y. 2004, \apjl, 616, L1

\bibitem[{{Hogerheijde} {et~al.}(1997){Hogerheijde}, {van Dishoeck}, {Blake},
  \& {van Langevelde}}]{hogerheijde:1997a}
{Hogerheijde}, M.~R., {van Dishoeck}, E.~F., {Blake}, G.~A., \& {van
  Langevelde}, H.~J. 1997, \apj, 489, 293

\bibitem[{{Hogerheijde} {et~al.}(1998){Hogerheijde}, {van Dishoeck}, {Blake},
  \& {van Langevelde}}]{hogerheijde:1998}
{Hogerheijde}, M.~R., {van Dishoeck}, E.~F., {Blake}, G.~A., \& {van
  Langevelde}, H.~J. 1998, \apj, 502, 315

\bibitem[{{Johnstone} {et~al.}(2004){Johnstone}, {Di Francesco}, \&
  {Kirk}}]{johnstone:2004}
{Johnstone}, D., {Di Francesco}, J., \& {Kirk}, H. 2004, \apjl, 611, L45

\bibitem[{{J{\o}rgensen} {et~al.}(2007){J{\o}rgensen}, {Bourke}, {Myers}, {Di
  Francesco}, {van Dishoeck}, {Lee}, {Ohashi}, {Sch{\"o}ier}, {Takakuwa},
  {Wilner}, \& {Zhang}}]{jorgensen:2007}
{J{\o}rgensen}, J.~K., {Bourke}, T.~L., {Myers}, P.~C., {et~al.} 2007, \apj,
  659, 479

\bibitem[{{J{\o}rgensen} {et~al.}(2005){J{\o}rgensen}, {Bourke}, {Myers},
  {Sch{\"o}ier}, {van Dishoeck}, \& {Wilner}}]{jorgensen:2005}
{J{\o}rgensen}, J.~K., {Bourke}, T.~L., {Myers}, P.~C., {et~al.} 2005, \apj,
  632, 973

\bibitem[{{J{\o}rgensen} {et~al.}(2004){J{\o}rgensen}, {Hogerheijde}, {van
  Dishoeck}, {Blake}, \& {Sch{\"o}ier}}]{jorgensen:2004}
{J{\o}rgensen}, J.~K., {Hogerheijde}, M.~R., {van Dishoeck}, E.~F., {Blake},
  G.~A., \& {Sch{\"o}ier}, F.~L. 2004, \aap, 413, 993

\bibitem[{{J{\o}rgensen} {et~al.}(2002){J{\o}rgensen}, {Sch{\"o}ier}, \& {van
  Dishoeck}}]{jorgensen:2002}
{J{\o}rgensen}, J.~K., {Sch{\"o}ier}, F.~L., \& {van Dishoeck}, E.~F. 2002,
  \aap, 389, 908

\bibitem[{{Lada}(1987)}]{lada:1987}
{Lada}, C.~J. 1987, in IAU Symp. 115: Star Forming Regions, ed. M.~{Peimbert}
  \& J.~{Jugaku}, 1--17

\bibitem[{{Lada} \& {Wilking}(1984)}]{lada:1984}
{Lada}, C.~J. \& {Wilking}, B.~A. 1984, \apj, 287, 610

\bibitem[{{Looney} {et~al.}(2000){Looney}, {Mundy}, \& {Welch}}]{looney:2000}
{Looney}, L.~W., {Mundy}, L.~G., \& {Welch}, W.~J. 2000, \apj, 529, 477

\bibitem[{{Ohashi} {et~al.}(1997){Ohashi}, {Hayashi}, {Ho}, \&
  {Momose}}]{ohashi:1997}
{Ohashi}, N., {Hayashi}, M., {Ho}, P.~T.~P., \& {Momose}, M. 1997, \apj, 475,
  211

\bibitem[{{Ossenkopf} \& {Henning}(1994)}]{ossenkopf:1994}
{Ossenkopf}, V. \& {Henning}, T. 1994, \aap, 291, 943

\bibitem[{{Palla} \& {Stahler}(1993)}]{palla:1993}
{Palla}, F. \& {Stahler}, S.~W. 1993, \apj, 418, 414

\bibitem[{{Pontoppidan} {et~al.}(2003){Pontoppidan}, {Fraser}, {Dartois},
  {Thi}, {van Dishoeck}, {Boogert}, {d'Hendecourt}, {Tielens}, \&
  {Bisschop}}]{pontoppidan:2003}
{Pontoppidan}, K.~M., {Fraser}, H.~J., {Dartois}, E., {et~al.} 2003, \aap, 408,
  981

\bibitem[{{Qi}(2005)}]{qi:2005}
{Qi}, C. 2005, The MIR Cookbook, The Submillimeter Array / Harvard-Smithsonian
  Center for Astrophysics (http://cfa-www.harvard.edu/~cqi/mircook.html)

\bibitem[{{Ridge} {et~al.}(2006){Ridge}, {Di Francesco}, {Kirk}, {Li},
  {Goodman}, {Alves}, {Arce}, {Borkin}, {Caselli}, {Foster}, {Heyer},
  {Johnstone}, {Kosslyn}, {Lombardi}, {Pineda}, {Schnee}, \&
  {Tafalla}}]{ridge:2006}
{Ridge}, N.~A., {Di Francesco}, J., {Kirk}, H., {et~al.} 2006, \aj, 131, 2921

\bibitem[{{Sault} {et~al.}(1995){Sault}, {Teuben}, \& {Wright}}]{sault:1995}
{Sault}, R.~J., {Teuben}, P.~J., \& {Wright}, M.~C.~H. 1995, in ASP Conf. Ser.
  77: Astronomical Data Analysis Software and Systems IV, 433

\bibitem[{{Shirley} {et~al.}(2002){Shirley}, {Evans}, \&
  {Rawlings}}]{shirley:2002}
{Shirley}, Y.~L., {Evans}, II, N.~J., \& {Rawlings}, J.~M.~C. 2002, \apj, 575,
  337

\bibitem[{{Ulrich}(1976)}]{ulrich:1976}
{Ulrich}, R.~K. 1976, \apj, 210, 377

\bibitem[{{Whitney} {et~al.}(2003){Whitney}, {Wood}, {Bjorkman}, \&
  {Wolff}}]{whitney:2003}
{Whitney}, B.~A., {Wood}, K., {Bjorkman}, J.~E., \& {Wolff}, M.~J. 2003, \apj,
  591, 1049

\bibitem[{{Wilking} {et~al.}(2005){Wilking}, {Meyer}, {Robinson}, \&
  {Greene}}]{wilking:2005}
{Wilking}, B.~A., {Meyer}, M.~R., {Robinson}, J.~G., \& {Greene}, T.~P. 2005,
  \aj, 130, 1733

\bibitem[{{Young} {et~al.}(2003){Young}, {Shirley}, {Evans}, \&
  {Rawlings}}]{young:2003}
{Young}, C.~H., {Shirley}, Y.~L., {Evans}, II, N.~J., \& {Rawlings}, J.~M.~C.
  2003, \apjs, 145, 111

\bibitem[{{Zinnecker} {et~al.}(1988){Zinnecker}, {Perrier}, \&
  {Chelli}}]{zinnecker:1988}
{Zinnecker}, H., {Perrier}, C., \& {Chelli}, A. 1988, in Proceedings of
  NOAO-ESO Conference on High-Resolution Imaging by Interferometry:
  Ground-Based Interferometry at Visible and Infrared Wavelengths, F. Merkle
  ed.

\end{thebibliography}

\end{document}